\documentclass[aps,prl,reprint,twocolumn]{revtex4-1}
\usepackage{graphics}
\usepackage{bbm}
\usepackage{bm}
\usepackage{dsfont}
\usepackage{amsmath,amssymb}
\usepackage{graphicx}
\usepackage[caption=false]{subfig}
\usepackage{ulem}
\usepackage{color}
\usepackage{xcolor}
\usepackage[uline]{hhtensor}

\begin{document}

\title{On the Geometry and Topology of Transformation Optics}%

\author{Yong-Liang Zhang$^1$, Li-Na Shi$^2$, Xian-Zi Dong$^1$, Fu-Zhou Gong$^3$,Zhen-Sheng Zhao$^1$ and Xuan-Ming Duan$^{4}$}%
\affiliation{$^1$Technical Institute of Physics and Chemistry, Chinese Academy of Science, Beijing 100190, China}
\affiliation{$^2$Institute of  Microelectronics, Chinese Academy of Sciences, Beijing 100029, China}
\affiliation{$^3$Institute of Applied Mathematics, Chinese Academy of Science, Beijing 100190, China}
\affiliation{$^4$Chongqing Institute of Green and Intelligent Technology, Chinese Academy of Science, Chongqing 400714, China}

\begin{abstract}
We generalize the geometrical model of transformation optics to Rieman-Cartan space with torsion by introducing topological defects in physical space. By relaxing the integrable condition, we show explicitly that the generalized equivalent medium are bi-anisotropic where the magnetoelectric coupling parameters emergent as the dislocation density. We also show the generation of orbital angular momentum of light. Our theory may open intriguing venues for controlling the vectorial degree of freedom of light with metamaterials.
\end{abstract}
\maketitle
\textit{Introduction.}---The analogy of electromagnetic waves in response to transparent media and the curved spacetime is a matter of extensive interest for both fundamental and technological reasons \cite{Skrotski:1957,Post,Landau:field,Lax:1976,Leonhardt:2006NJP,Visser}. Notably, the discovery of transformation optics sheds new lights on this pursuit \cite{Leonhardt:2006,Milton:2006}. Analogous to the general relativistic description of gravity, transformation optics roots on the nontrivial Riemannian structure of the physical space. It enables a variety of unprecedented electromagnetic phenomena such as invisibility cloaks \cite{Smith:2006}, optical illusion \cite{Lai:2009}, imaging \cite{Smith:2010}, light harvesting \cite{Pendry:2012} to nonlocal plasmonics\cite{Maier:2012}. Physically, the permittivity and permeability specified by the coordinate transformation determine the local optical length and fix the causal structure of the space-time. In this sense, the metric induced equivalent medium can be regard as a perfect geometrical optics material, where a light ray follows the null geodesic regardless of its polarization state \cite{Leonhardt:2006NJP,Leonhardt:2006}.

Though great progress has been made in the paradigm of transformation optics and it's several extensions \cite{Mccall:2011,Thompson:2011}, some fundamental aspects on the treatment of the transversal nature of electromagnetic fields, such as bianisotropy, polarization state and singular phase structure, have remained elusive. The covariant formalism of linear classical electrodynamics \cite{Post,Landau:field,Thompson:2011,Hehl:2008} and constraint of electromagnetic duality symmetry \cite{Molina-Terriza:2013} allow more generic bi-anisotropic constitutive laws including magnetoelectric coupling terms \cite{Lindell:1994}. The magnetoelectric coupling effect represents the first order of weak spatial dispersion which results from the nonlocal response of polarization and magnetization due to the optically non-negligible size and distance, or the chiral configuration of the inclusions. From the viewpoint of topology, it is inherently related with the singular phase structure in the near field.

In this letter, we show that the generic transformation media corresponding to the curved vacuum space are bi-anisotropic where the anisotropic magnetoelectric coupling terms come from the torsion tensor of the physical space. Torsion was first introduced by Cartan as the asymmetric part of the connection, $T^{\gamma}_{\ \alpha\beta}=(\Gamma^{\gamma}_{\ \alpha\beta}-\Gamma^{\gamma}_{\ \beta\alpha})/2$, to incorporate the intrinsic angular momentum of elementary particles into general relativity \cite{Hehl:1976}. It plays a fundamental role in the gauge theories of gravitation \cite{Hehl:2013}, though the observable effects is still not experimentally detected due to the weak gravitational coupling. Here, we show that the physical space transformed from a Riemannian background material is a general Riemann-Cartan space endowed with metric and torsion. In our scheme, torsion is defined in terms the spatial curls of the coordinate transformation and intimately relate with the presence of line topological defects in the physical space \cite{Kondo:1964,Kroner:1960,Kleinert:2008}. As physical effects, we demonstrate the generation of both spin and orbital angular momentum of light by a toy model of optical cosmic strings. The torsion induced physical effects are topological protected against continuous deformations and yet flexible for the same reason, allowing for tunability without loss of functionality.

\textit{Generalized transformation optics.}---We begin by considering an arbitrary coordinate transformation from a virtual simple connected flat space, denoted by $\mathbf{r}(x,y,z):=\{x^a\}$ with Roman indices, to a transformed physical space, denoted by coordinates $\mathbf{r'}(x',y',z'):=\{x^{\alpha}\}$ with Greek indices. For simplicity, we restrict ourselves to the case of Euclidean transformation in three dimensional Cartesian system. However, our formalism should be applicable to four dimensional Minkowski space-time directly. The two coordinate systems are related with the Pfaffian form \cite{Schouten}
\begin{equation}\label{Pfaff}
\vartheta^{\alpha}=e_a^{\ \alpha}dx^a
\end{equation}
with the transformation Jacobian matrix $e_a^{\ \alpha}=\partial x^{\alpha}/\partial x^{a}$. Hereafter, Einstein summation convention over repeated indices is applied. If the one-form (\ref{Pfaff}) is non-integrable, the frame $\{\partial_\alpha\}$ constitutes a non-coordinate base. We can introduce the object of anholonomity $\Omega^{\delta}_{\ \beta\gamma}=e^{\ \delta}_a(\partial_\beta e^a_{ \ \gamma}-\partial_\gamma e^a_{\ \beta})$ which is the structure constant of the basis vector $[e_\alpha,e_\beta]=\Omega^\gamma_{\ \alpha\beta}e_\gamma$ \cite{Schouten}. Assuming the time harmonic fields are $\propto \mathrm{exp}(-i\omega t)$ with frequency $\omega$, the spatially covariant curl Maxwell equations in free space can be expressed with anholonomic coordinates,
\begin{eqnarray}
\label{Maxwell}
\begin{split}
\epsilon^{\alpha\beta\gamma}\left[\partial_\beta E_\gamma+\Omega^{\delta}_{\ \beta\gamma}E_\delta\right]-i\omega{B}^\alpha=0, \\
\epsilon^{\alpha\beta\gamma}\left[\partial_\beta H_\gamma+\Omega^{\delta}_{\ \beta\gamma}H_\delta\right]+i\omega{D}^\alpha=0
\end{split}
\end{eqnarray}
where $\epsilon^{\alpha\beta\gamma}$ is the completely antisymmetric Levi-Civita symbol, and  in right-handed coordinate system it is a tensor density $\epsilon^{\alpha\beta\gamma}=e^{\alpha\beta\gamma}/\sqrt{g}$, where $g$ is the determinant of the metric, $e^{\alpha\beta\gamma}=0, \pm 1$ is the permutation symbol. Lower the upper indices, the curl equations (\ref{Maxwell}) can be reorganized to:
\begin{eqnarray}\label{anholonomic Maxwell}
\begin{split}
e^{\alpha\beta\gamma}\partial_\beta E_\gamma-i\omega \bar{B}^\alpha=0, \\
e^{\alpha\beta\gamma}\partial_\beta H_\gamma+i\omega \bar{D}^\alpha=0
\end{split}
\end{eqnarray}
Eqs. (\ref{anholonomic Maxwell}) resemble the macroscopic Maxwell equations in dielectric with the constitutive equations,
\begin{eqnarray}\label{constitutive relations}
\begin{split}
\bar D^\alpha=\varepsilon_0\varepsilon^{\alpha\beta}E_\beta-i\frac{1}{c}\kappa^{\alpha\beta} H_\beta, \\
\bar B^\alpha=i\frac{1}{c}\kappa^{\alpha\beta} E_\beta+\mu_0\mu^{\alpha\beta}H_\beta
\end{split}
\end{eqnarray}
where the material parameters are given by
\begin{eqnarray}
\label{constitutive}
\begin{split}
\varepsilon^{\alpha\beta}=\mu^{\alpha\beta}=\sqrt{g}g^{\alpha\beta},  \\
\kappa^{\alpha\beta}=\frac{\lambda_0}{4\pi}e^{\alpha\gamma\delta}\Omega^{\beta}_{\ \gamma\delta} \
\end{split}
\end{eqnarray}
 here $g^{\alpha\beta}=e^{\ \alpha}_{a}e^{\ \beta}_a$ is the contravariant metric of the physical space, and $\lambda_0=2\pi c/\omega$ is the vacuum wavelength. According to the formal invariance of Maxwell equations, we interpret (\ref{anholonomic Maxwell}) as the macroscopic Maxwell equations in nontrivial transformed media in Cartesian coordinates. The constitutive relations (\ref{constitutive relations}) is a standard formula of material equations for linear bianisotropic media. Because $\Omega^{\gamma}_{\ \alpha\beta}$ is antisymmtric with respect to the lower indices, the magnetoelectric coupling tensors have nine independent components. Equations (\ref{constitutive relations},\ref{constitutive}) which generalize the constitutive relations of the transformed media from anisotropic to bi-anisotropic materials with magnetoelectric coupling terms, are the fundamental results of this work.

The essential feature of the present work is that the magnetoelectric coupling stems from geometry due to coordinate transformation but instead of the bi-anisotropic background \cite{Mackay:2008,Kong:2009} or moving dielectrics \cite{Skrotski:1957,Post,Leonhardt:2006NJP,Landau:field}. From the view point of differential geometry, the geometric information of the Riemannian physical space is encoded in the Jacobian matrix through the nontrivial metric $\mathbf{g}=e \mathrm{\eta} e^T$, where $\eta=\mathrm{diag}\{1,1,1\}$ is the metric of flat space, and the metric compatible Levi-Civita connection $\tilde{\Gamma}^{\gamma}_{\ \alpha\beta}$ which is symmetric on the lower indices. In anholonomic coordinates, the affine connection is given by
$\Gamma^{\gamma}_{\ \alpha\beta}=\tilde{\Gamma}^{\gamma}_{\ \alpha\beta}-\Omega^{\gamma}_{\ \alpha\beta}+g_{\beta\delta}g^{\gamma\rho}\Omega^{\delta}_{\ \alpha\rho}+g_{\alpha\delta}g^{\gamma\rho}\Omega^{\delta}_{\ \beta\rho}$ \cite{Schouten}. In our case, we start from a flat Euclidean space with a trivial connection and obtain an asymmetric connection due to the object of anholonomity. As a result, the torsion tensor is then obtained
\begin{eqnarray}
T^{\gamma}_{\ \alpha\beta}=-\frac{1}{2}\Omega^{\gamma}_{\ \alpha\beta}
\end{eqnarray}
Consequently, the generalized physical space in transformation optics is a Riemann-Cartan space with nonvanishing torsion which act as the MCP terms for the transformed media.

\begin{figure}[htbp]
\centerline{\scalebox{0.45}{\includegraphics{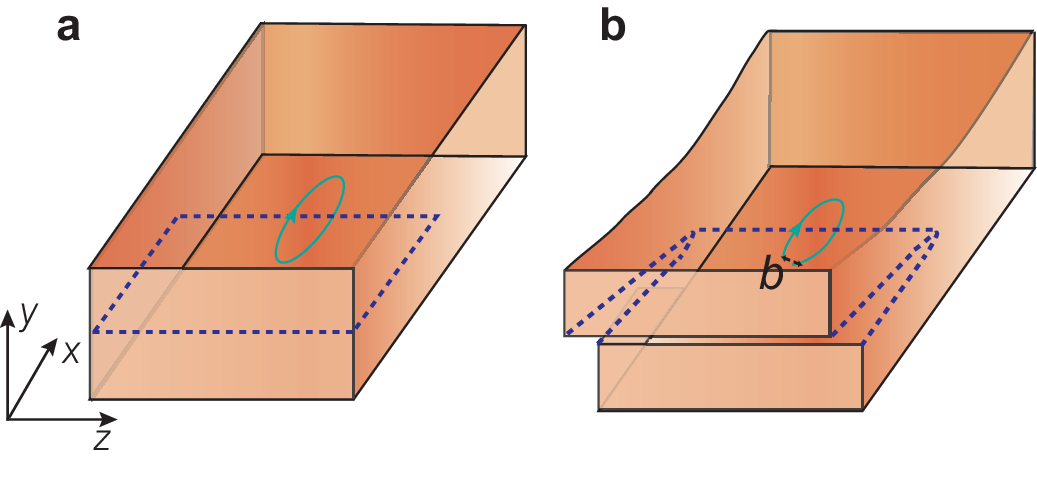}}}
\caption{\label{fig1}Schematic illustration of nonholomonic coordinate transformation between a perfect domain (Fig. 1a) to a distorted one (Fig. 1b) by Volterra construction. The failure of the closed loop (sea green) indicates a line defect along $z$-axis.}
\end{figure}

The emergent magnetoelectric coupling results from the homogeneity breaking of the translational symmetry of the coordinate transformation. In fact, torsion vanishes if the displacement vector $\mathbf{u}(\mathbf{r})=\mathbf{r}'-\mathbf{r}$ is a smooth single valued function, which corresponds to a elastic deformation preserving the topology between the virtual and physical spaces. In this case, the one-form (\ref{Pfaff}) is integrable and $\vartheta^\alpha=dx^\alpha$ is a total differential. If $\vartheta^\alpha$ is non integrable, the magnetoelectric coupling emerges due to the violation of Frobenius integrability condition $\partial_{\mu}e^a_{\ \nu}=\partial_{\nu}e^a_{\ \mu}$ \cite{Schouten,Kleinert:2008}, which is usually ignored in the relevant literature \cite{Leonhardt:2006NJP,Leonhardt:2006}. The resulting incompatible transformation is a plastic deformation which changes the topology of the physical space. In this case, $\vartheta^\alpha$ is not a total differential and the displacement vector is a multivalued function. This implies a jump discontinuity for $\mathbf{u}(\mathbf r)$ along any closed contour $\mathbf{b}=\oint d\mathbf{u}$. Using Stokes theorem, the vector $\mathbf{b}$ can be expressed as
\begin{equation}\label{guage field}
b^a=\int_{S} \Omega^a_{\ \alpha\beta}d\sigma^{\alpha\beta}
\end{equation}
where $d\sigma^{\alpha\beta}=\epsilon^{\alpha\beta\gamma}n_{\gamma}d\sigma$ is the differential area, $n_\gamma$ is the unit normal vector, and the field $\Omega^a_{\ \alpha\beta}$ related the object of anholonomity by $\Omega^\gamma_{\ \alpha\beta}=e^{\ \gamma}_{a}\Omega^a_{\ \alpha\beta}$.  Because affine connection determine the parallel transport of vectors, torsion describes  the broken of infinitesimal parallogram in physical space, and relates the Euclidean translations in transformation optics. For anholonomic transformations, Eq. (\ref{guage field}) indicates that there is a change in topology between the virtual and physical spaces, and the object of anholonomity measures the failure of closing for infinitesimal parallelogram, which is analogous to the Burgers circuit describing  dislocations in crystalline solids. With the presence of a discrete dislocation, the physical space becomes multiple connected. A pure screw or edge dislocation representing the closed failure can be created by a standard Volterra's cut and glue procedure.  It is obvious that the displacement vector, which varies smoothly everywhere except on the defect core, is a multivalued function when the defect line is encompassed. This allows us to distinguish different loops encircling the defect in physical space and so maybe classified by the fundamental group $\pi_1(S^1)=\mathbb{Z}$. In the continuum limit, torsion is proportional to the density of topological charge. It should be noted that $\mathbf{b}$ is a topological invariant in the sense that it is independent of the particular choice of integral paths provided it encompasses fixed defects.

The intrinsic length scale of torsion also provides an interpretation to the factor $\lambda_0/4\pi$ for the magnetoelectric coupling terms. Similar factor is also found in the formal analysis of the frequency domain Maxwell's equations in Einstein-Cartan spacetime \cite{Horsley:2011}, where the isotropic chiral parameter was determined by a pure axial vector form of the dual torsion preserving gauge invariance and principle of semi-minimal coupling \cite{Hojman:1978}. But, the torsion in our model comes from the non-vanishing object of anholonomity which describes the rotation of the local orthogonal frame.  And the nine independent components of torsion give rise to the anisotropic magnetoelectric coupling parameter.

An interesting question is how torsion affect the symmetry of generalized transformation optics: (i) By construction, $\kappa$ and $\gamma$ are odd under parity $\mathcal{P}$ and time reversal $\mathcal{T}$ with $[\mathcal{P},\mathcal{T}](\mathbf{E},\mathbf{H})=(-\mathbf{E},\mathbf{H})$. Thus they are  pseudo tensors to retain the form of Maxwell equations \cite{Lindell:1994,Horsley:2011}; (ii) The constitutive relations (4,5) satisfy with constraint due to the electromagnetic duality symmetry \cite{Molina-Terriza:2013}. The source free Maxwell's equations with constitutive relation (\ref{constitutive relations}) are invariant with respect to the continuous duality rotation
\begin{eqnarray}
\label{DC}
\left[\begin{array}{ccc}
\mathbf{E}_\theta \\
\mathbf{H}_\theta
\end{array}\right]=\left[\begin{array}{ccc}
cos\theta & -sin\theta \\
sin\theta & \ cos\theta
\end{array}\right]
\left[\begin{array}{ccc}
\mathbf{E} \\
\mathbf{H}
\end{array}\right]
\end{eqnarray}
Because helicity operator is the generator of electromagnetic duality symmetry \cite{Molina-Terriza:2013}, the duality invariance leads to the conservation of helicity. As a result, different helicity states (circular polarization) do not couple with each other when light waves propagate in an ideally generic transformation medium.

\textit{The topological aspects.}---Let us establish the connection between the formalism we have set up in previous section with the nontrivial topology of the physical space by the integral forms of Maxwell's equations \cite{Lax:1976}. Consider the Faraday's law for a spatially uniform monochromatic waves reads $\oint_{\Gamma} \mathbf{E}\cdot d\mathbf{r}=i\omega\int_{S} \mathbf{B}ds$. Under the anholonomic transformation, the closed surface $S$ is mapped to a surface $S'$  with boundary $\Gamma'$ with additional Burgers vector $\mathbf{b}$. In physical space, a straightforward integration for the electric field yields
\begin{equation}
\label{Faraday}
\oint_{\Gamma'}\mathbf{E}'\cdot d\mathbf{r}'=i\omega\int_{S'} \mathbf{B'}\cdot d\mathbf{\sigma}'+\mathbf{b}\cdot\mathbf{E'}
\end{equation}
A treatment for Amp\`{e}re's circuital law yields a similar expression. In (\ref{Faraday}), the presence of dislocation density $\mathbf{b}$  indicates the topological nature of MCP, since it is invariant under a continuous deformation for the integral loop and it's value depends upon a physical quantity in a region outside the domain of integration. It should be noted that torsion should has the same spatial components as the electric or the magnetic field to manifest the physical effects.

Although the polarization state of light is invariant in generalized transformation media, it changes when light pass through the boundary between two distinct media. For simplicity we consider a boundary between vacuum and a transformed medium. It is known that the polarization state of monochromatic light is closely related to Lipkin's Zilch \cite{Lipkin:1964} $\rho_\chi=(\epsilon_0\mathbf{E}\cdot\nabla\times\mathbf{E}+\mu_0^{-1}\mathbf{H}\cdot\nabla\times\mathbf{H})/2$, which measuring the optical chirality of the optical field \cite{Bliokh:2011}. Under holonomic coordinate transformation $\mathbf{r}\rightarrow\mathbf{r}'$, this pseudoscalar density transforms as $\rho_{\chi}'(\mathbf{r}')=J\rho_{\chi}(\mathbf{r})$ with $J=\mathrm{det}\left(\frac{\partial \mathbf{r}}{\partial \mathbf{r}'}\right)$. However, there is an additional term in anholonomic coordinates
\begin{eqnarray}
\rho_{\chi}'=J\rho_{\chi}+\rho_{T}
\end{eqnarray}
where $\rho_T=\epsilon^{\alpha\beta\gamma}e^a_\alpha\partial_\beta e^b_\gamma\left(\varepsilon_0E_aE_b+\mu_0^{-1}H_aH_b\right)/2$ is a torsional analog of the Chern-Simons like term in three dimensional space \cite{Hehl:1995}. $\rho_T$ transforms as a scalar density under a gauge transformation $e^a_{\ \mu}\rightarrow e^a_{\ \mu}+\partial_{\mu}\Lambda$ with $\Lambda$ a continuous function. It is interesting that the gauge transformation here can be interpreted as a continuous deformation upon the coordinate transformation. Therefore, the integral $\int d^3x\rho_\chi$ is invariant under coordinate transformation. This invariance indicates that the integral of density of optical chirality over transformed space is a topological invariant depend only on the topology of the physical space. This can also be verified by the fact that the tensor indices are contracted with the Levi-civita symbol $\epsilon^{\mu\nu\lambda}$ instead of the metric. Consequently, an ideal torsional medium will turn a linearly polarized plane wave to an elliptical polarized one.

We can also relate torsion with optical vortices carrying orbital angular momentum, since vortices are also known as optical phase dislocations \cite{Berry:1973}. Let us consider a general scalar plane wave in virtual space $\psi=\rho \mathrm{exp}(i\chi(\mathbf{\mathbf{r}}))$, with intensity $\rho^2$ and phase $\chi(\mathbf{r})=\mathbf{k}\cdot\mathbf{r}$. Under the action of anholonomic transformation, the phase in physical space becomes $\chi'=\mathbf{k}\cdot(\mathbf{r}+\mathbf{u})$. By a contour integral of phase on a closed circuit $\mathcal{C}$ around the origin, we can obtain the winding number $N$,
\begin{equation}\label{winding number}
N=(2\pi)^{-1}\oint_{\Gamma'} \nabla\chi'\cdot d\bm{r}'=\mathbf{b}/2\pi
\end{equation}
From Eq. (\ref{winding number}), it is evident that an anholonomic mapping at macroscopic length scale converts a plane wave into a dislocated wavefront beam with the topological strength $N$. Specifically, a pure screw dislocation when $\bm k=(0,0,k)$ correspond to a optical vortex with an exp$(il\theta)$ azimuthal phase dependence, where $l=N/k$. More generic singular optical fields can be produced by distributed pure or mixed optical dislocations \cite{Berry:1973}.


\textit{A toy model.}---To elucidate the application of our theory, let us introduce a toy model that describes the spatial part in the exterior of a straight cosmic string \cite{Letelier:1993}
\begin{equation}\label{metric}
r'=r,\ \theta'=\alpha\theta, \ z'=z+\frac{b}{2\pi}\theta
\end{equation}
where $r=\sqrt{x^2+y^2}$, $\theta=\mathrm{atan}(y/x)$, and $\mathbf{b}=(0,0,b)$ is the Burgers vector. Due to the singular nature along $z$ axis, $T^z$ is the only non-vanishing component of torsion associated with this line element, $T^z\propto b\delta^2(r)$ \cite{Jackiw:1990,Letelier:1993, Kleinert:2008}. To avoid the complex boundaries, we consider the coordinate transformation (12) mapped the imaginary upper surface of the cylinder with twisted shape to a planar surface of a cylinder in physical space, and the corresponding material parameters are given by

\begin{eqnarray*}
\varepsilon=\mu=\alpha^{-3}\left[
\begin{array}{ccc}
sin^2\theta+\alpha^2cos^2\theta & sin\theta cos\theta [\alpha^2-1] & \gamma sin\theta \\
& cos^2\theta+\alpha^2sin^2 & -\gamma cos\theta \\
&  & \alpha^2+\gamma^2 \\
\end{array}
\right]
\end{eqnarray*}
and
\begin{equation}\label{material parameters}
\displaystyle
\kappa^{zz}=-\gamma^{zz}=\frac{\lambda_0}{2\pi}\delta^{(2)}(r)=\lim_{n\rightarrow\infty}\frac{\lambda_0bn}{2\pi^{3/2}}e^{-n^2r^2}
\end{equation}
Here, we have used the Gaussian sequence for planar Dirac $\delta$-function $\delta(r)=\lim\limits_{n\to\infty}\frac{n}{\sqrt{\pi}}e^{-n^2r^2}$. 
However, a metamaterial made from 2D arrays of subwavelength defects are chiral metamaterials with emerging magnetoelectric coupling \cite{Giessen:2012}.

\textit{Summary and outlook.}---We have established a generic framework for transformation optics with torsion by breaking the topology of the physical space. This generalization could open new horizons for the technological applications where the spin and orbital angular momentum of electromagnetic waves are relevant. The combination of torsional effects into current framework allows for the full control of electromagnetic waves with coordinate transformation well beyond the intuitive picture of ray optics. Our formulism here not only provides a theoretical foundation to understand the fundamental principle of transformation optics but also promises a model system for simulating of classical and quantum physics in curved torsional space in optical laboratory.

\begin{acknowledgments}
This work was financially supported by the National Basic Research Program (2010CB934102); Natural Science Foundation of China (NSFC, No. 60907019, No. 61077028, No. 50973126); The International Cooperation Program under Grant (No. 2008DFA02050, No. 2010DFA01180) of Most of China.
\end{acknowledgments}

\end{document}